# Fabrication and characterization of suspended microstructures of ultrananocrystalline diamond

C. Pachiu [1], T. Sandu [1,2], C.Tibeica [1,2], L. M. Veca [1], R. Popa [1],
M. Popescu [1], R. Gavrila [1], C. Popov [3], V. Avramescu [1]

[1] National Institute for Research and Development in Microtechnologies - IMT Bucharest,
126A Erou Iancu Nicolae, Bucharest, Romania
cristina.pachiu@imt.ro

[2] Research Centre for Integrated Systems, Nanotechnologies, and Carbon Based Materials - IMT Bucharest
sandu.titus@imt.ro

[3] Institute of Nanostructure Technologies and Analytics, University of Kassel, Heinrich-Plett-Strasse 40, 34132 Kassel, Germany

**Abstract.** The fabrication of various suspended microstructures made from ultrananocrystalline diamond was performed without sacrificial layer. Arrays of bridges and cantilevers of various dimensions were successfully fabricated by a well-controlled process. Mechanical characterizations and finite element calculations allowed us to estimate both the Young's modulus as well as the residual stress built in the ultrananocrystalline diamond film.

## 1. Introduction and preliminary results

Diamond, in its single-crystal form, possesses rather exceptional properties [1, 2] that are attractive for many applications working in extreme conditions such as environments with high temperature or/and corrosive chemicals, high-speed/ high-power switches, etc. Its mechanical properties (hardness of 100 GPa, Young's modulus of 1200 GPa, very good wear resistance and low friction) together with its high chemical stability and biocompatibility make diamond and diamond-based micro-electromechanical systems (MEMS) good for environmentally friendly applications in information technology, automotive and aerospace industries, and medicine [3, 4].

These impressive properties of single-crystal diamond are expected to be found to a large extent into nanocrystalline diamond (NCD) and ultrananocrystalline diamond

(UNCD). NCD and UNCD are two-phase systems composed of nanometer-sized diamond grains and amorphous carbon that fills the inter-crystalline space. The NCD films contain grains with sizes between tens of nm and one µm. On the other hand, the UNCD films have grains with sizes varying from 2 to 10 nm [5].

Being grown much easier than the single-crystalline films [6, 7] and displaying smooth surfaces, uniform thicknesses, and mechanical properties close to those of single-crystal diamond, the UNCD films are ideal for MEMS applications. [8]. The fabrication of MEMS devices is a multi-step process of deposition, photo-lithographic patterning, and selective etching of a multi-layered thin film [9]. In this fabrication process a sacrificial silicon dioxide layer is used. The sacrificial layer is deposited/grown with the scope of supporting the subsequent device layer and later is removed by a selective isotropic etching in order to release a free-standing MEMS device. Thus, the structure of the multi-layered thin film consists of a device layer at the top, a sacrificial layer in the middle which is supported by a substrate at the bottom. During the final stages of removing the sacrificial layers stiction problems may appear. Stiction is the unintentional adhesion of otherwise freestanding structures to the substrate. There are several ways to deal with the stiction problem. One of them is to roughen the surface to reduce the contact area [10]. There are also other ways to solve the problem like the changing of wetting properties of the solution-substrate interface [11], the use of a supercritical drying [12], etc.

In this work we follow a different approach by omitting completely the sacrificial layer to fabricate suspended UNCD microstructures. In other words, the multi-layered thin film is just a micrometer-sized UNCD film directly deposited on the silicon substrate. Then, while maintaining the integrity of the UNCD microstructures, we remove sufficient material from underneath of the UNCD device by a deep isotropic etching process to obtain suspended microstructures. The main motivation for not using a sacrificial layer is the fact that a direct growth of UNCD films on $SiO_2$ is not straightforward and it needs the addition of Ar in plasma for successful growth of UNCD film on sacrificial layer $SiO_2$ [13]. Thus, we were able to fabricate suspended structures like bridges and cantilevers without the use of a sacrificial layer. Moreover, it is well-known that during the fabrication process, the mechanical stress, in general, and the residual stress, in particular, may play an important role for MEMS structures especially in the final stages of fabrication after the release from the substrate [4]. We also evaluated the residual compressive stress of the UNCD film deposited on the silicon substrate by measuring the Young's modulus and using its value in FEM (Finite Element Method) mechanical simulations. The work has the following structure: section 2 presents the process fabrication of suspended UNCD based microstructures; section 3 is dedicated to the morphological and mechanical characterization and to discussions; in the last section 4 we conclude the work.

## 2. UNCD device fabrication

In this work we used UNCD films grown by microwave plasma enhanced chemical vapor deposition (MW-PECVD) on 3-inch (100) Si wafers [7]. The film pattering was obtained first by fabricating an aluminum (Al) mask directly on the UNCD film. Then, the UNCD microstructures were obtained with plasma assisted etching of (a) the UNCD film in oxygen atmosphere and (b) the silicon substrate in sulfur-hexafluoride ($SF_6$) (Reactive Ion Etching - RIE with EtchLab SI 220, Sentech Instruments). Both etching processes were performed in the presence of the metallic mask. Some of these results showing released microstructures were presented in a recent conference paper [14].

In the following we present the details of the full fabrication process. The fabrication starts with the deposition of the metal layer. Prior to metal deposition, the UNCD surface was treated in a cleaning hot piranha solution ($H_2SO_4$, $H_2O_2$ 3:1 at 150$^o$C for 5 min, followed by washing in deionized water and drying in nitrogen) to modify the surface into oxygen- and hydroxyl-terminated one for a better adhesion.

The metal deposition was made directly on the UNCD film. Different materials, photoresists, and metals with different etch rates and selectivity can be used in mixture of oxygen and sulfur hexafluoride ($SF_6$) RIE, but we have chosen aluminum as a mask in UNCD etching process because it has a significantly lower etch rate and a stronger resistance to etching than a photoresist [15, 16]. Thus, a 500 nm thick Al film was sputtered onto the UNCD layer (AUTO 500, BOC Edwards) (Figure 1.a, b).

In a second step, the photoresist was deposited by coating to configure the mask on Al layer (Figure 1.c). Using the classical photolithography technique, the metallic mask was patterned with the designed geometry (Figure 1.d). To obtain micron-wide features, a 1.5 μm thick positive photoresist was used (HPR-504, Fujifilm), spin-coated at 3000 rpm, followed by the post-baking at 100º C for 30 min, and developed in 15-30 seconds with HPRD-437 developer. The mask contained the following targeted microstructures: arrays of micro-cantilevers and bridges.

The process of metal mask fabrication was continued with the wet chemical etching of the metal layer through the patterned photoresist (3: 3: 1: 1 $H_3PO_4$: $HNO_3$: $CH_3COOH$: $H_2O$ at 40° C for 2 min) [17] at 40ºC for about 2 min, followed by the removal of the remaining photoresist.

In this way, we obtained the metallic protective mask that is needed to fabricate the UNCD - based microstuctures such as micro-cantilevers and bridges.

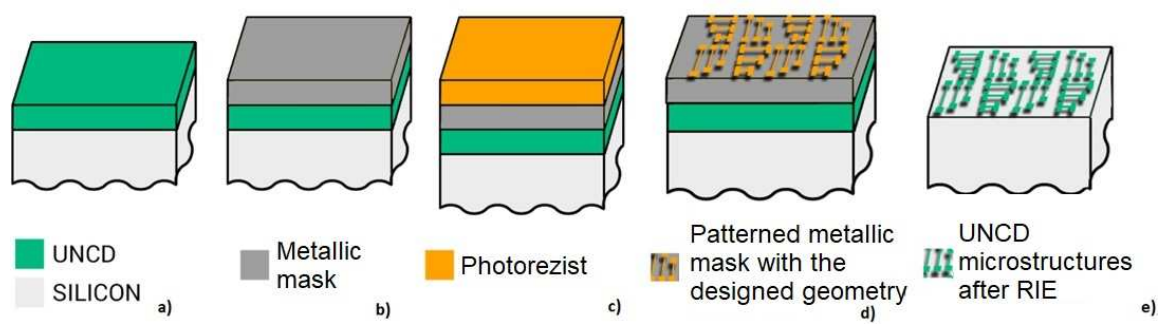

**Figure 1.** Fabrication flow of UNCD microstructures: (a) the UNCD film deposited on Si; (b) the deposition of the metal layer used to fabricate the protecting mask; (c) the photoresist deposition; (d) the lithographic process pattering the metallic layer; (e) the processes of etching the UNCD film and the Si substrate to release the microstructures.

Using the metallic mask, the UNCD film was processed with selective etching by reactive ion etching (RIE) in oxygen atmosphere, with 450 Watts RF power under a pressure of 15 Pa. The etching rate of the UNCD film was approximately 65 nm/min using an $O_2$ gas flow rate of 50 sccm for 10 min.

In order to release the patterned microstructures, the process was continued with two recipes for isotropic etching of Si substrate (Figure 1.e). The first etching recipe was as follows: mixture of oxygen and sulfur hexafluoride ($SF_6$) RIE for 10 min, with an $O_2$ gas flow rate of 50 sccm and $SF_6$ gas flow rate of 10 sccm. The etching rate was approximately 90 nm/min. The second etching recipe was performed at the same RF power and gas pressure using not more than $SF_6$, with a gas flow rate of 100 sccm for 10 min. The above patterning processes were successfully tested and consequently arrays of bridges and cantilevers were fabricated using them.

## 3. Results and discussions

### 3.1. Characterization of UNCD - based microstructures

Various suspended microstructures were fabricated and inspected by high-resolution electron microscopy in order to evaluate the technological efficiency of the processes described in the previous section. Detailed examination of the fabricated structures was conducted by SEM (Scanning Electron Microscopy), employing Nova NanoSEM 630 (former FEI Company, today Thermo Fisher Scientific, USA).

One of the challenges in the fabrication of UNCD suspended microstructures was the robustness of Al mask with respect to the etching agents used in the etching processes. During the UNCD etching process, oxygen removes also Al creating $Al_2O_3$ debris that are taken away from the mask and moved into the regions where the UNCD film is supposedly etched away. The debris acts as micro-masks for the UNCD material in the etching region, hence the UNCD etching is incomplete [18]. Initially, we thought that, to get rid of remaining UNCD material, it is better to keep oxygen as an etching agent together with $SF_6$ during the Si etching process (the first recipe described in the previous section).

However, it turns out that not removing the photoresist in the process of fabricating the Al mask together with the use of $SF_6$ alone as an etching agent for Si substrate lead to a successful fabrication of suspended microstructures (Figure 4).

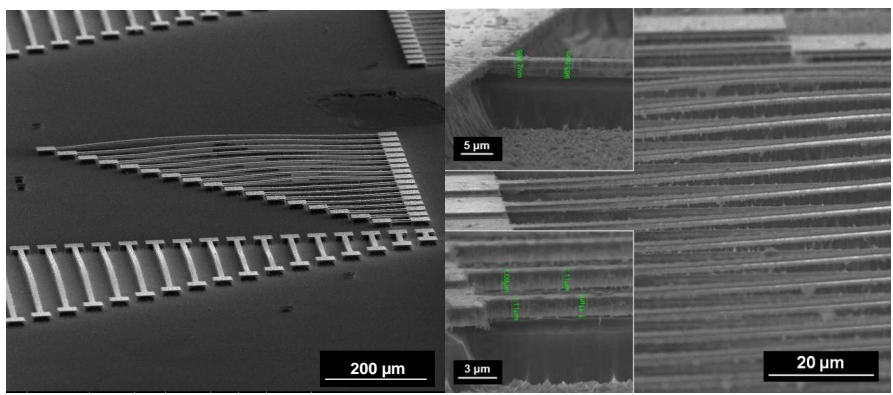

**Figure 2.** SEM micrographs of arrays of bridges fabricated from UNCD films in the oxygen and sulfur hexafluoride mixture RIE process. Various degrees of buckling indicating a compressive stress can be observed. In the zoom images the presence of debris can be seen.

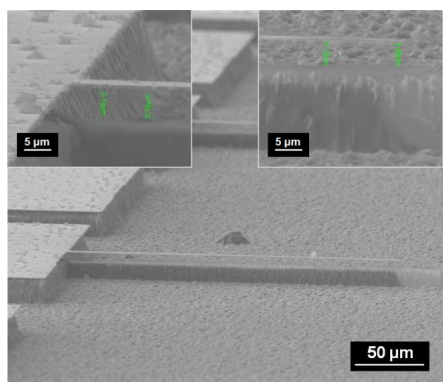

**Figure 3.** SEM micrographs of cantilevers fabricated from UNCD films in the oxygen and sulfur hexafluoride mixture RIE process. In the zoom images the presence of debris can be seen.

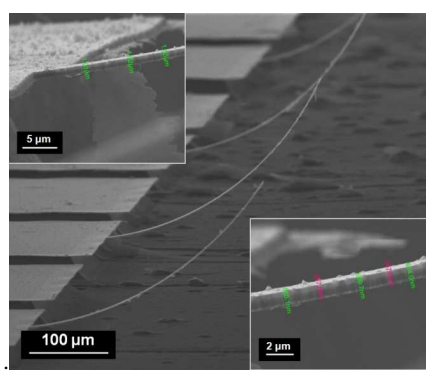

**Figure 4.** SEM micrographs of cantilevers fabricated from UNCD films by etching the Si substrate with sulfur hexafluoride only. As it can be seen in the zoom images the debris effects were minimized. The images were taken at an intermediate phase of the process, when the Al mask was not removed from the microstructures. Al from the mask creates an additional stress that warps the cantilevers.

From Figures 2 and 3 one can see that the bridges and the cantilevers are completely released from substrate. The effect of adding oxygen in the process of Si etching can be seen in zoomed images presented in Figures 2 and 3. Nevertheless, this process creates debris acting as micro-masks which make a rough and uneven substrate underneath of fabricated microstructures. On the contrary, when only $SF_6$ is used to etch Si, the substrate underneath of microstructures looks much better, hence the effect of the debris is minimized (Figure 4). We notice here that the images shown in Fig. 4 were taken before removing the Al mask. The upward bending of cantilevers indicates an induced stress by the Al mask.

### 3.2. Young modulus measurements of UNCD films

The Young's modulus of the UNCD films was measured by depth-sensing indentation (Nano Indentation) technique employing a G200 Nano Indenter (Keysight Technologies, former Agilent Technologies). The indenter tip used in the measurements was a three-sided pyramidal Berkovich diamond tip.

The patented Continuous Stiffness Measurement (CSM[TM]) method was employed. It allowed the assessment of the mechanical behavior of the material composing the film independently from that of the underlying substrate. In the CSM method, an oscillating force considerably smaller than the nominal load is superimposed on the quasi-static indentation. The sample response is then analyzed by a lock-in amplifier and the value of the Young's modulus (E) is calculated by the instrument software as a function of the indentation depth $d$, using the well-established Oliver and Pharr method [19].

Ten indentation tests were performed on different sites of the film surface. A characteristic curve of E as a function of depth was then computed by averaging the individual measurements. The Young's modulus ($E$) of the film was derived as the plateau value of the indentation modulus $E = f(d)$ curve (Figure 5). The estimated values of Young modulus of the UNCD film were about 300-315 GPa, which are smaller than those reported in the literature with values between 500 and 1000 GPa [13, 18, 20]. This can be attributed to the presence of the grain boundary material. Further tests with various film thicknesses are needed in order to rule out the effect of the substrate on our measurements. Yet, this value is much higher than the Young's modulus of Si.

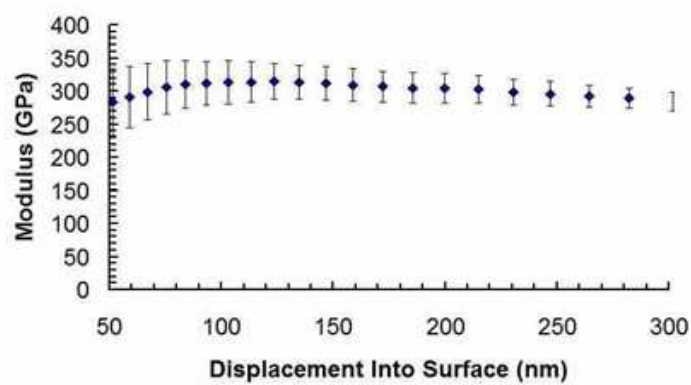

**Figure 5.** The graph of recorded Young modulus as a function of indentation depth into the film surface.

### 3.3. Stress estimation of UNCD films from analysis of FEM calculations of fabricated microstructures

Series of double clamped beams of different lengths (in the range 300 - 510 µm) were characterized by WLI (White Light Interferometry - FOGALE 3D). Their profiles show the first buckling mode (Figure 6), which means a compressive residual stress existing in the UNCD film.

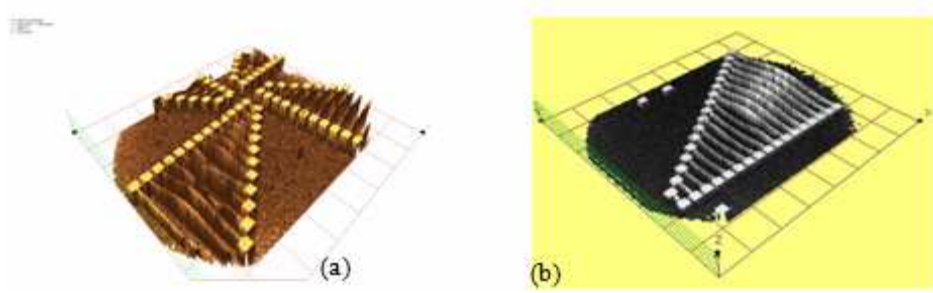

**Figure 6.** 3D WLI images of the bridge arrays; (a) the reconstructed profiles of four constant-width groups (widths of: 4.5, 6, 7.5, and 9 μm, respectively); (b) the reconstructed profile of a constant-width group which has bridges of increasing length (length range: 300 -510 μm).

For each of the bridges the buckling amplitudes, $s$, were measured and tabulated in Table 1.

**Table 1.** Measured amplitudes ($s$) of the first buckling mode.

| L [μm] | 300 | 330 | 360 | 390 | 420 | 450 | 480 | 510 |
|---|---|---|---|---|---|---|---|---|
| s [μm] | 5.92 | 6.55 | 7.06 | 7.65 | 8.19 | 8.76 | 9.32 | 9.91 |

For guiding purposes, the assessment of the magnitude of the residual stress from buckling was initially considered with an analytical formula. The compressive residual stress is given analytically by the following expression [21, 22]:

$$\sigma_0 = \frac{E\pi^2}{L^2}\left(\frac{s^2}{4} + \frac{t^2}{3}\right) \quad (1)$$

where $E$ is the Young's modulus of the material, $t$ the beam's thickness, $s$ the out-of-plane displacement amplitude of the first buckling mode, and $L$ the length of the beam.

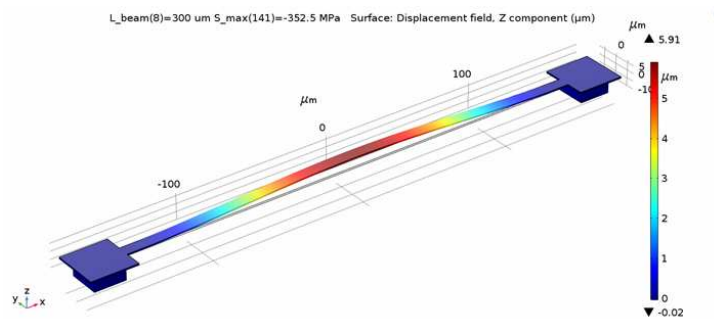

**Figure 7.** A FEM simulated bridge (L=300 μm) that includes the bond pads and the silicon pillars, showing the displacement in the first buckling mode due to the compressive stress

However, the analytical model is an idealization of the experimental situation, hence FEM (finite element method)-based simulations would be more appropriate. We performed FEM calculations using COMSOL Multiphysics software to find out the value of the residual stress. In this case we assumed a uniform bi-axial compressive stress in the plane of the UNCD layer, and a zero-gradient stress along the thickness, namely $\sigma_{xx}(z) = \sigma_{yy}(z) = \sigma_0$. The FEM model was built using the actual geometry of the fabricated structures and the same material properties as used in the analytical model. In order to find the stress magnitude that creates the beams shapes, a variable parametric simulation was conducted, and the results were compared to the measured values of bucking mode amplitude for each simulated beam.

The dimensions of the microstructures and the material properties used in simulations were those obtained from the design parameters and from measurements. Thus, the thickness of the UNCD film, $t$, was set to 1 µm, and the Young's modulus, $E$, to 300 MPa, whilst the values of $L$ and $s$ were taken from Table 1. A typical FEM calculation of the bucking due the residual stress is presented in Figure 7, where bond pads and Si pillars were considered. In Figure 8 we show the results of the film stress calculated by both the analytical formula, Eq. (1), and by FEM simulations.

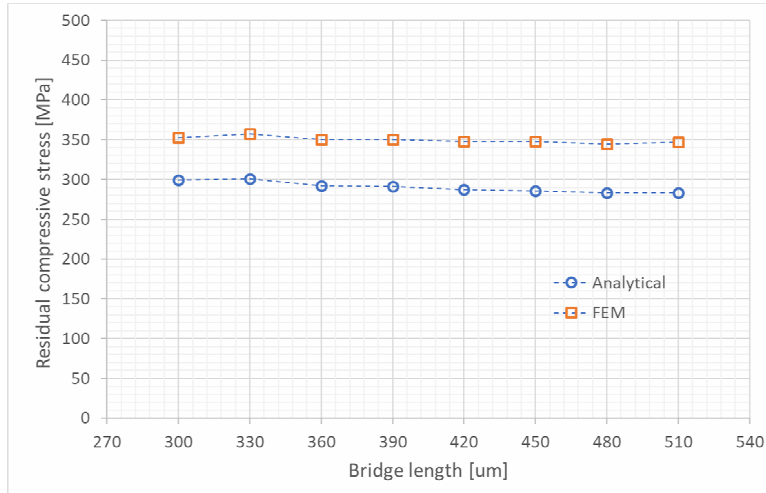

**Figure 8.** Analytical versus FEM calculations of stress values for different beam lengths.

The analytical model gives an average value of the compressive stress of about 290 MPa, while in the FEM calculations the value of stress is 350 MPa. The difference between these two approaches is of approximately 20% but we must bear in mind that the FEM model was built to reproduce various details of the real object. The FEM model includes the bond pads of the bridges, the undercut, and the silicon pillars. Thus, a series of effects such as the non-ideal clamping (Figure 9) are not considered in the ideal beam model (the analytical approach) but are taken into account in the FEM model. Also, in the FEM analysis the geometric nonlinearities were considered, contrary to the analytical model where they are ignored. The value we have calculated for the residual stress in the UNCD film is within the bounds of the residual stress found in the literature, where a stress of a few hundreds of MPa can be encountered in UNCD films [13, 18, 20].

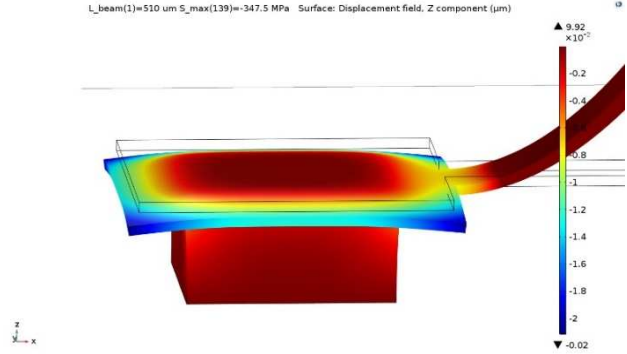

**Figure 9.** Image showing the non-ideal clamping of the beam's end in the FEM model. Displacements are exaggerated by a factor of 100 in order to show the effect.

The buckling profiles obtained by WLI were used to check the correctness of the analytical and FEM models. The z-displacement of the beam within the analytical model of the first buckling mode is described by the following sinusoidal function [21, 22]:

$$z(x) = \frac{s}{2}\left(1 + \cos\frac{2\pi x}{L}\right), \qquad (2)$$

where $x$ is the coordinate in the $xy$ - plane of the beam. We must notice here that only the bucking amplitude $s$ is involved in Eq. (2), but not the residual stress. As it can be seen from Figures 10 and 11, the analytical and FEM simulated shapes are almost identical to the WLI profile.

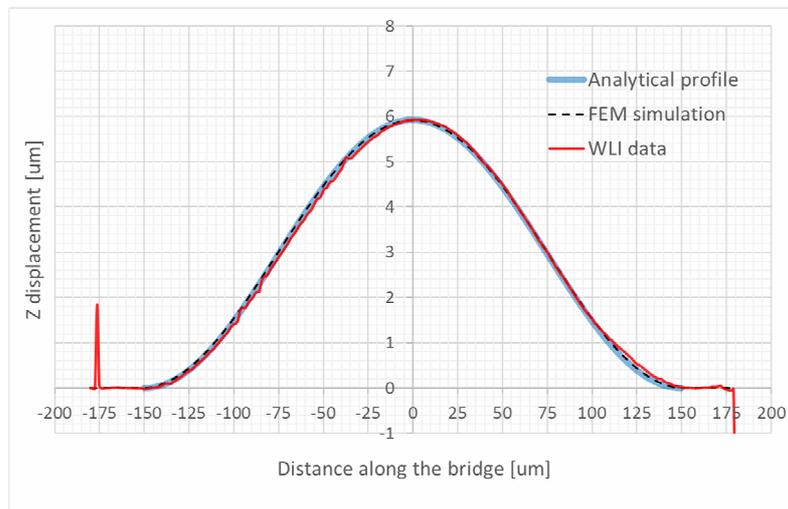

**Figure 10.** Comparison of WLI data, the FEM-based simulation, and the analytical profile of the first buckling mode of a 300 μm long beam.

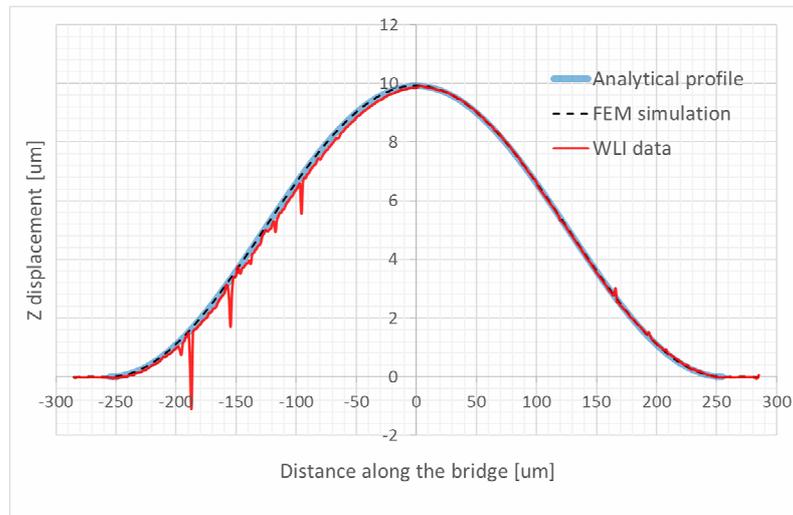

**Figure 11.** WLI data against the FEM-based simulation and against the analytical profile of the first buckling mode of a 500 μm long beam.

## 4. Conclusions

In the present work we studied the fabrication of different suspended microstructures from UNCD films grown on Si substrate. The fabrication was performed without sacrificial layer by isotropic etching of the substrate. Through a well-controlled multi-step process, we fabricated arrays of bridges and cantilevers of various dimensions. The bridges were buckled showing a residual stress built in the UNCD film. In order to estimate the residual stress, the Young's modulus was measured. The measurements gave us a value of Young's modulus that ranges between 300 and 315 GPa. The knowledge of Young's modulus as well as the measurements of buckling profiles by White Light Interferometry allowed the determination of the residual stress by finite element calculations performed with COMSOL software. Finite element calculations provided us a stress value of 320 MPa, which is 20% higher than the value given by the analytical formula of an ideal beam model.

**Acknowledgements.** This research was supported by ANCSI, CORE - Programme No. PN 16 32 02 01/2016: "Carbon nanostructures – experimental and applicative investigations".